\documentclass[10pt,authoryear]{article}

\usepackage[authoryear]{natbib}
\usepackage{lineno}
\usepackage{amsmath}
\usepackage{mathtools}
\usepackage{ae,aecompl}
\usepackage[]{graphicx}
\usepackage{amsfonts}
\usepackage{amssymb}
\usepackage{epstopdf}
\usepackage{epsfig}
\usepackage{float}
\def\lsim{\lower.5ex\hbox{$\; \buildrel < \over \sim \;$}}
\def\gsim{\lower.5ex\hbox{$\; \buildrel > \over \sim \;$}}

\providecommand{\keywords}[1]{\textit{Keywords:} #1}

\begin{document}

\title{Influence of the geometric configuration of accretion flow on the black hole spin dependence of relativistic acoustic geometry}

%\author[Tarafdar, Ananda \& Das]{
%Pratik Tarafdar,$^{1}$\thanks{E-mail: pratik.tarafdar@bose.res.in}
%Deepika B. Ananda,$^{2}$\thanks{E-mail: deepika@camk.edu.pl}
%Tapas K. Das,$^{3}$\thanks{E-mail: tapas@hri.res.in}
%\\
%$^{1}$S. N. Bose National Centre For Basic Sciences, Block-JD, Sector III, Salt Lake, Kolkata, India.\\
%$^{2}$Nicolaus Copernicus Astronomical Centre of the Polish Academy of Sciences, Warsaw, Poland.\\
%$^{3}$Harish-Chandra Research Institute, Chhatnag Road, Jhunsi, Allahabad, India.\\
%}

\author{Pratik Tarafdar\\
	S. N. Bose National Centre For Basic Sciences\\ 
	Block-JD, Sector III, Salt Lake, Kolkata, India.\\
	pratik.tarafdar@bose.res.in
	\and
	Tapas Kumar Das\\
	Harish-Chandra Research Institute\\
	Chhatnag Road, Jhunsi, Allahabad, India.\\
	tapas@hri.res.in}
\date{}

\maketitle

\begin{abstract}
\noindent
Linear perturbation of general relativistic accretion of low angular momentum hydrodynamic fluid onto a Kerr black hole leads to the formation 
of curved acoustic geometry embedded within the background flow. Characteristic features of such sonic geometry depend on the black hole spin. 
Such dependence can be probed by studying the correlation of the acoustic surface gravity $\kappa$ with the Kerr parameter $a$. The $\kappa - a$ 
relationship further gets influenced by the geometric configuration of the accretion flow structure. In this work, such influence has been studied 
for multitransonic shocked accretion where linear perturbation of general relativistic flow profile leads to the formation of two analogue 
black hole type horizons formed at the sonic points and one analogue white hole type horizon which is formed at the shock location producing 
divergent acoustic surface gravity. Dependence of the $\kappa - a$ relationship on the geometric configuration has also been studied for 
monotransonic accretion, over the entire span of the Kerr parameter including retrograde flow. For accreting astrophysical black holes, 
the present work thus investigates how the salient features of the embedded relativistic sonic geometry may be determined not only by 
the background space-time, but also by the flow configuration of the embedding matter.
\end{abstract}

%\begin{keywords}
%accretion, accretion discs -- black hole physics -- hydrodynamics -- shock waves -- gravitation
%\end{keywords}

\section{Introduction}
Linear perturbation of inhomogenous inviscid transonic fluid leads to the formation of black hole like 
acoustic space time, characterized by an acoustic metric possessing some horizon-embedded features of the fluid flow 
\cite{unruh81prl,visser98cqg,blv05lrr,us07}.
Recent works 
\cite{abd06cqg,dbd07jcap,pmdc12cqg}
demonstrate the emergence of such acoustic geometry and corresponding multiple acoustic horizons - one white hole 
type and two black hole type - within the general relativistic accretion of matter onto astrophysical black holes, 
either of the Schwarzschild or of the Kerr type. In these works, however, only a particular type of flow 
geometry (either flow with constant disc thickness or flow in hydrostatic equilibrium along the vertical direction) has been considered 
to study such emergent gravity phenomena. Since there might be three different geometric configurations with which 
matter may accrete onto astrophysical black holes 
(see, e.g. section $4$ of \cite{bcdn14cqg}, for the details of such configurations), 
it is necessary to observe how the characteristic features of the embedded acoustic metric get influenced by the 
flow geometry of the accreting matter. 
\cite{td15ijmpd}
have studied the polytropic as well as the isothermal relativistic accretion onto a non-rotating black hole for the 
three different geometric configurations, viz.- flow with constant thickness, flow with quasi-spherical configuration 
and flow in hydrostatic equilibrium along the vertical direction (`constant height flow' (CH), `conical flow' (CF) and 
`vertical equilibrium flow' (VE), respectively, hereafter). This has been accomplished to understand how the value of 
the acoustic surface gravity, which is one of the primary quantifiers of the analogue gravity phenomena, gets influenced 
by the matter geometry as well as with the equation of state describing the infalling matter. \\
It is, however, to be noted that most of the astrophysical black holes are believed to be of Kerr type, i.e., they possess 
non-zero values of the black hole spin parameter $(a)$ 
\cite{mrfng09apj,dwrb10mnras,kmtnm10mnras,tnm10apj,ziolkowski10,bggnps11a,daly11mnras,mr11mnras,mndgkoprs11cqg,nckp11mnras,
rbltmrnf12aip,tm12mnras,brenneman13,dcppv13apj,mtb13s,fpwmkrd14mnras,hlz14arxiv,jbs14arxiv,nt14arxiv,sbdr14arxiv}. 
Hence, it is necessary to understand how the black hole spin dependence of the prominent features of the acoustic geometry 
gets influenced by the matter geometry. In our present work, which may be considered as the sequel of 
\cite{td15ijmpd}, 
we address this issue. In this connection, it is to be noted that purely from astrophysical viewpoint, 
\cite{tbnd17na} 
have investigated the integral stationary accretion solutions onto a Kerr black hole to understand the multi-transonic 
structure of such flow and the formation of the stationary shock transitions. In their work, the parameter space spanned 
by various initial boundary conditions (governing the flow) has been constructed to show how the stationary shocks 
form for various flow geometries as well as for various equations of state. In summary, the multi-transonic properties of 
general relativistic accretion (adiabatic and isothermal) of matter in the Kerr metric has been thoroughly analyzed in the 
aforementioned work to understand how the geometric configuration of accretion flow influences the black hole spin dependence 
of the multi-transonic flow profile for such accretion. \\
Our present work will take directions from the results obtained by 
\cite{tbnd17na}. 
We would like to study the variation of the acoustic surface gravity with the black hole spin for multi-transonic shocked 
accretion solutions in three different matter geometries. We thus identify the regions of the parameter space responsible for 
shock formation using the formalism introduced in 
\cite{tbnd17na}. 
Once we identify such parameters, we shall use the respective boundary conditions to calculate the value of the analogue 
surface gravity in terms of the accretion variables defined at the critical points (see subsequent sections for detailed description 
of such procedure). We then study the variation of the values of such acoustic surface gravity with the black hole spin. We study 
such variation for three different geometric configurations of matter to understand how the $\kappa - a$ dependence gets influenced 
by the matter geometry. We perform such analysis for both polytropic as well as for isothermal accretion.  

\section{Outline of the solution scheme}
We consider low angular momentum inviscid accretion onto a Kerr black hole along the equatorial plane. The space-time is described 
by Boyer-Lindquist line element of the following form (normalized for $G=c=M_{BH}=1$), where $G$, $c$ and $M_{BH}$ are the universal 
gravitational constant, velocity of light in vacuum and mass of the black hole respectively. 
\begin{equation}
ds^2=g_{\mu \nu}dx^\mu dx^\nu=-\frac{r^2\Delta}{A}dt^2+\frac{A}{r^2}(d\phi-\omega dt)^2+\frac{r^2}{\Delta}dr^2+dz^2
\label{eqn1}
\end{equation} 
where, $\Delta=r^2-2r+a^2$, $A=r^4+r^2a^2+2ra^2$, $\omega=\frac{2ar}{A}$, $a$ being the Kerr parameter. \\
Following usual procedure 
% (references)
, the acoustic surface gravity $\kappa$ may be expressed as, 
\begin{equation}
\kappa=\left[\kappa_0\left(\frac{du}{dr}-\frac{dc_s}{dr}\right)\right]_{r=r_h}
\label{eqn2}
\end{equation}
where $r_h$ denotes the location of the acoustic horizon, which actually is the location of the sonic point for a transonic flow, 
$u$ represents the dynamical radial advective velocity defined on the equatorial plane, 
$c_s$ is the corresponding speed of sound, and
\begin{equation}
\kappa_0=\frac{\left(a^2+(r-2) r\right)\sqrt{\left(a^2 (r+2)+r^3-\lambda ^2 (r-2)-4 a \lambda\right)}}{\left(1-c_s^2\right) \left(a^2 (r+2)-2 a \lambda +r^3\right)\sqrt{r}}.
\label{eqn3}
\end{equation}

As is obvious from the above expression, for a certain pre-defined set of initial boundary conditions defined by 
the specific flow energy ${\cal E}$, specific angular momentum $\lambda$, the adiabatic index $\gamma$ and the Kerr parameter $a$, 
one has to use the critical point analysis to find out the critical points and thence, the sonic points by integrating the stationary 
flow solutions. Once the sonic point is obtained, the value of the advective velocity $u$, the sound speed $c_s$ and 
the respective space gradients of $u$ and $c_s$ will be calculated at the sonic points. 
Such sonic points are actually the acoustic horizons becasue the flow makes a transition from subsonic to supersonic state 
at those points. Once the set of values of the location of the sonic points $(r_s)$ and $\left[u,c_s,\frac{du}{dr},\frac{dc_s}{dr}\right]_{r_s}$ 
are obtained, $\kappa$ can be calculated for polytropic accretion for a set of initial boundary conditions prescribed by 
$\left[{\cal E},\lambda,\gamma,a\right]$. For isothermal accretion, the initial boundary conditions are specified by 
$\left[\lambda,T,a\right]$, where $T$ is the bulk flow (ion) temperature. Since isothermal sound speed is position independent, 
the corresponding value of the acoustic surface gravity can be calculated by knowing the location of the isothermal sonic point $r_s$ and 
$\left[u,c_s,\frac{du}{dr}\right]_{r_s}$, for a fixed set of values of $\left[\lambda,T,a\right]$. \cite{tbnd17na}(sections 2-6)
% (Reference) 
describes how to calculate $\left(r_s,\left[u,c_s,\frac{du}{dr},\frac{dc_s}{dr}\right]_{r_s}\right)$ for a particular set of 
$\left[{\cal E},\lambda,\gamma,a\right]$ for polytropic accretion for three different geometric configurations of the flow. 
Similar calculations are available for isothermal accretion as well (see, e.g. section 9-13 of \cite{tbnd17na}). \\

For certain values of $\left[{\cal E},\lambda,\gamma,a\right]$ and $\left[T,\lambda,a\right]$, stationary integral solutions 
corresponding to the adiabatic as well as to isothermal flow becomes multi-transonic. That is, the transition from subsonic to 
supersonic regime takes place twice, leading to the formation of two black hole like acoustic horizons. Flow lines between two 
such horizons are connected through a stationary shock. At the shock location, an acoustic white hole is formed where the 
acoustic surface gravity becomes infinite. In principle, what happens is as follows: \\
At a large distance from the gravitational horizon of the astrophysical black hole, accreting matter remains subsonic. 
Such subsonic matter encounters an outer sonic point and becomes supersonic. Such supersonic flow may then encounter a
stationary shock (depending on whether certain criteria, the general relativistic Rankine-Hugoniot condition for the polytropic flow, 
for example, are satisfied) and becomes subsonic, The shock transition is a discontinuous transition. $\left[u,c_s\right]$ 
changes discontinuously at the shock location. Hence, $\left[\frac{du}{dr},\frac{dc_s}{dr}\right]$ diverges at the shock and 
hence $\kappa$ diverges at the white hole as well, whereas the values of $\kappa$ at the two acoustic black holes, one located at the outer 
sonic point and the other located at the inner sonic point, remain finite. Section 5 of 
\cite{pmdc12cqg} 
presents the exemplary methodology for the calculation of $\kappa$ for a typical multi-transonic shocked accretion for flow in hydrostatic 
equilibrium along the vertical direction. \\
As mentioned in the introduction section that our main goal in this project is to study how the $\kappa - a$ dependence gets influenced 
by the geometrical configuration of matter. To accomplish this task, one thus needs to have idea about how the shock formation phenomenon 
is guided by the spin parameter $a$. Also since we need to compare $\kappa - a$ plot for three different matter geometries, it is required 
to obtain the aforementioned shock parameter space for all three geometries and then concentrate on the mutually overlapping region. \\
Following the methodology developed in 
\cite{tbnd17na}, 
we thus construct shock forming parameter space (spanned by $\lambda$ and $a$ keeping ${\cal E}$and $\gamma$ of the flow constant for 
adiabatic accretion, and by keeping $T$ of the flow constant for isothermal accretion) for the different flow geometries and then identify 
the common region of overlap. From the identified region, we shall choose the range of $a$ corresponding to a fixed set of 
$\left[{\cal E},\lambda,\gamma\right]$ for adiabatic accretion and $\left[T,\lambda\right]$ for isothermal accretion. For those values 
of $a$, we shall calculate the locations of sonic points $r_s$, the shock location $r_{sh}$ and $\left[u,c_s,\frac{du}{dr},\frac{dc_s}{dr}\right]_{r_s}$ 
for the adiabatic flow and $\left[u,c_s,\frac{du}{dr}\right]_{r_s}$ for the isothermal flow. The values of $r_s$, $u$, $c_s$, $\frac{du}{dr}$ and 
$\frac{dc_s}{dr}$ for various values of $a$ will then be substituted in the respective expression of $\kappa$ as obtained in eqn.(\ref{eqn2}), 
and $\kappa - a$ will be obtained for three different geometric configurations of matter.

\section{Dependence of $\kappa$ on $a$ for adiabatic accretion}

\begin{figure}[h!]
\centering
\begin{tabular}{cc}
\includegraphics[width=0.5\linewidth]{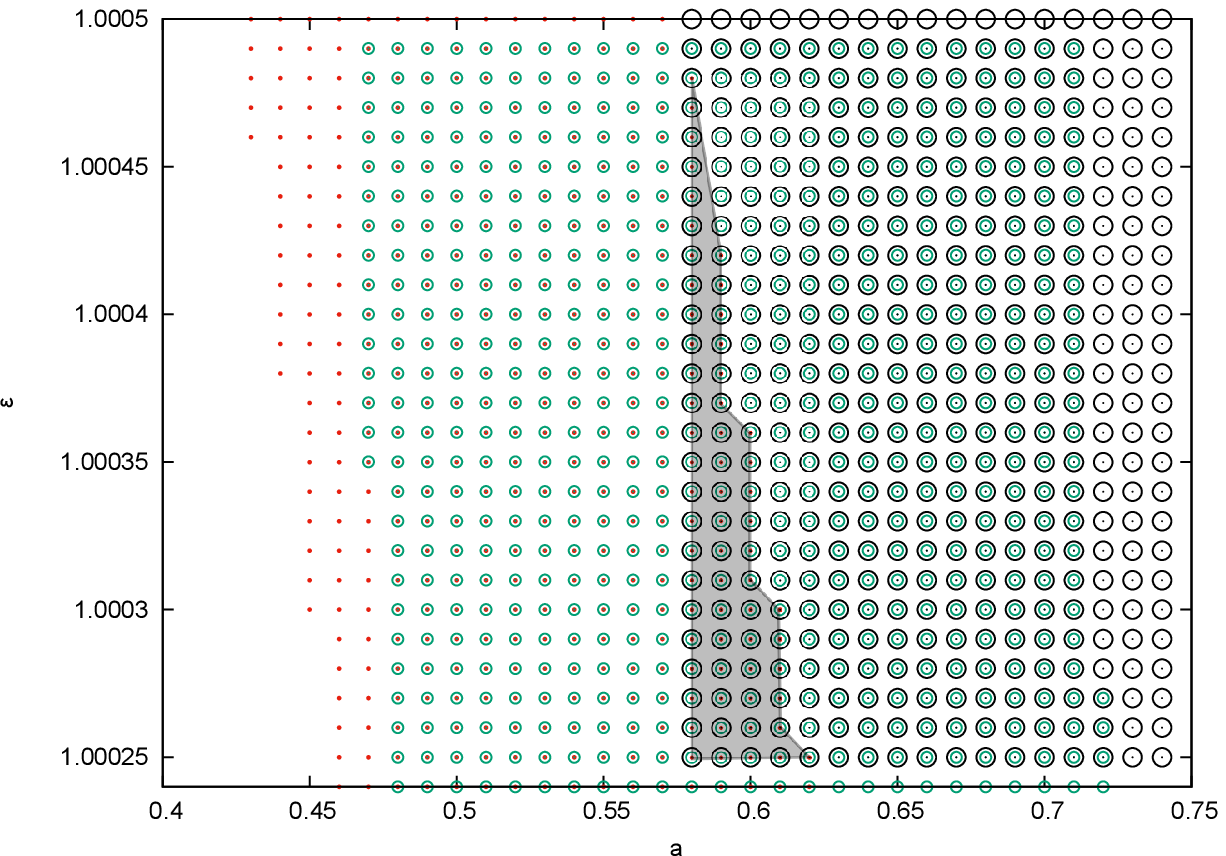} &
\includegraphics[width=0.5\linewidth]{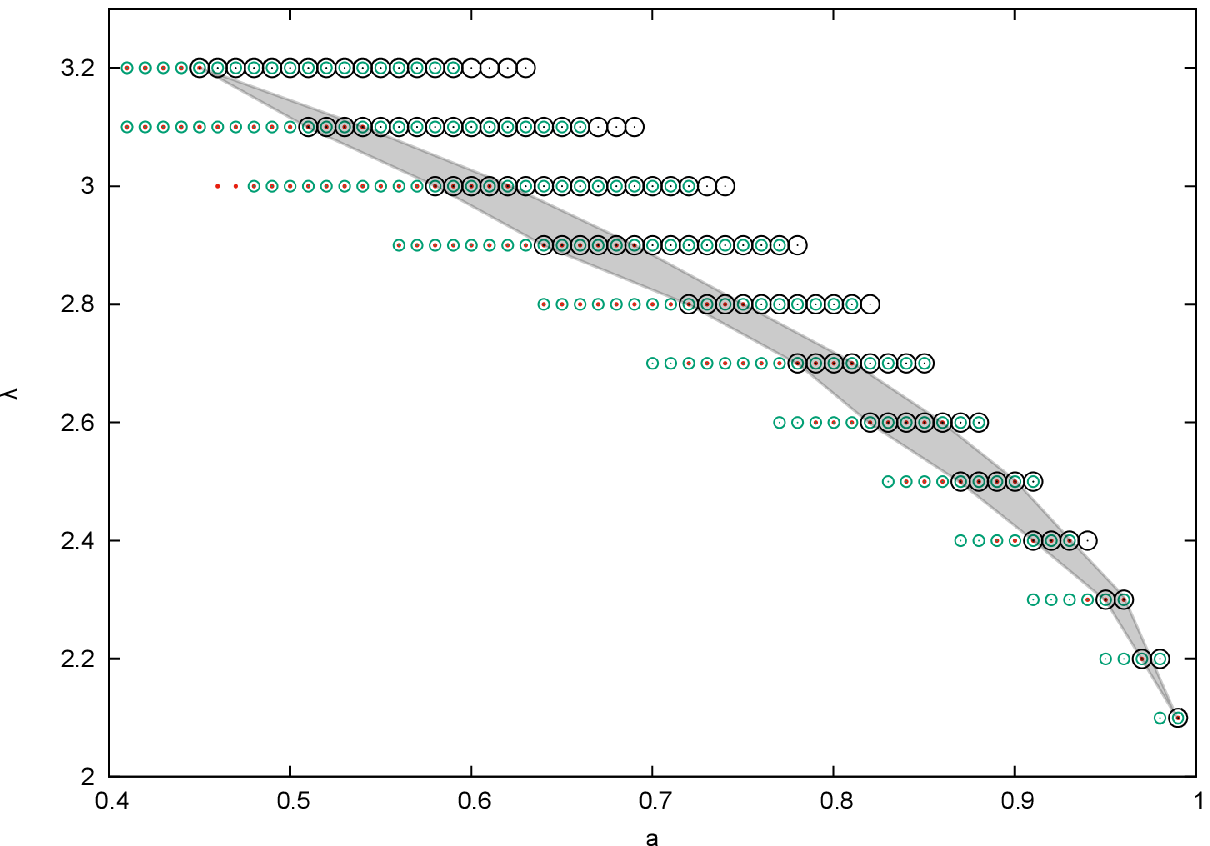}
\end{tabular}
\caption{${\cal E} - a$ ($\lambda=3.0$) and $\lambda - a$ (${\cal E}=1.00025$) plots for $\gamma=4/3$, showing the common shock overlap regions for 
the three disc geometries. Green circles, red dots and black circles represent CF, VE and CH flows respectively.}
\label{fig1}
\end{figure}

In fig. 1, we plot the parameter space region spanned by $\left[{\cal E},\lambda\right]$ and $a$ for which shock forms in all three 
geometric configurations of accreting matter for a fixed set of values of the remaining parameters ($\gamma=\frac{4}{3}$ and 
$\left[\lambda=3.0\right]$ for ${\cal E} - a$ diagram, $\left[{\cal E}=1.00025\right]$ for $\lambda - a$ diagram). One has to 
understand that we are interested in studying the variation of $\kappa$ with the black hole spin. It is thus necessary that out of 
all parameters determining the flow, the parameters other than the spin have to be kept constant and $a$ has to be varied to calculate 
$r_s$ and $\left[u,c_s,\frac{du}{dr},\frac{dc_s}{dr}\right]_{r_s}$ accordingly for all matter geometries. Hence, which parameter out of 
the rest is plotted along the y-axis is somewhat immaterial. A common region of shock formation for three different geometries (the shaded 
regions in fig. 1) can be 
demonstrated by plotting any of the other parameters against $a$. In the present picture, we have chosen such values of the fixed parameters 
so as to enhance the range of $a$ for which the common shock forming region forms. Similar figure can be produced by plotting $\gamma$ vs. $a$ 
as well. \\
Note that the multitransonic flow possesses two sonic points, $r_s^{in}$ (inner one, closer to the black hole event horizon) and $r_s^{out}$ 
(outer one, far away from the event horizon). Thus, we have two different values of the acoustic surface gravity, 
$\kappa^{in}$ defined at the inner sonic point, and $\kappa^{out}$ defined at the outer sonic point. At the shock, the accretion variables change 
discontinuously and therefore the space derivatives of such variables diverge. Hence, the acoustic surface gravity at the shock location $\kappa^{sh}$ 
cannot be computed. 

\begin{figure}[h!]
\centering
\begin{tabular}{cc}
\includegraphics[width=0.5\linewidth]{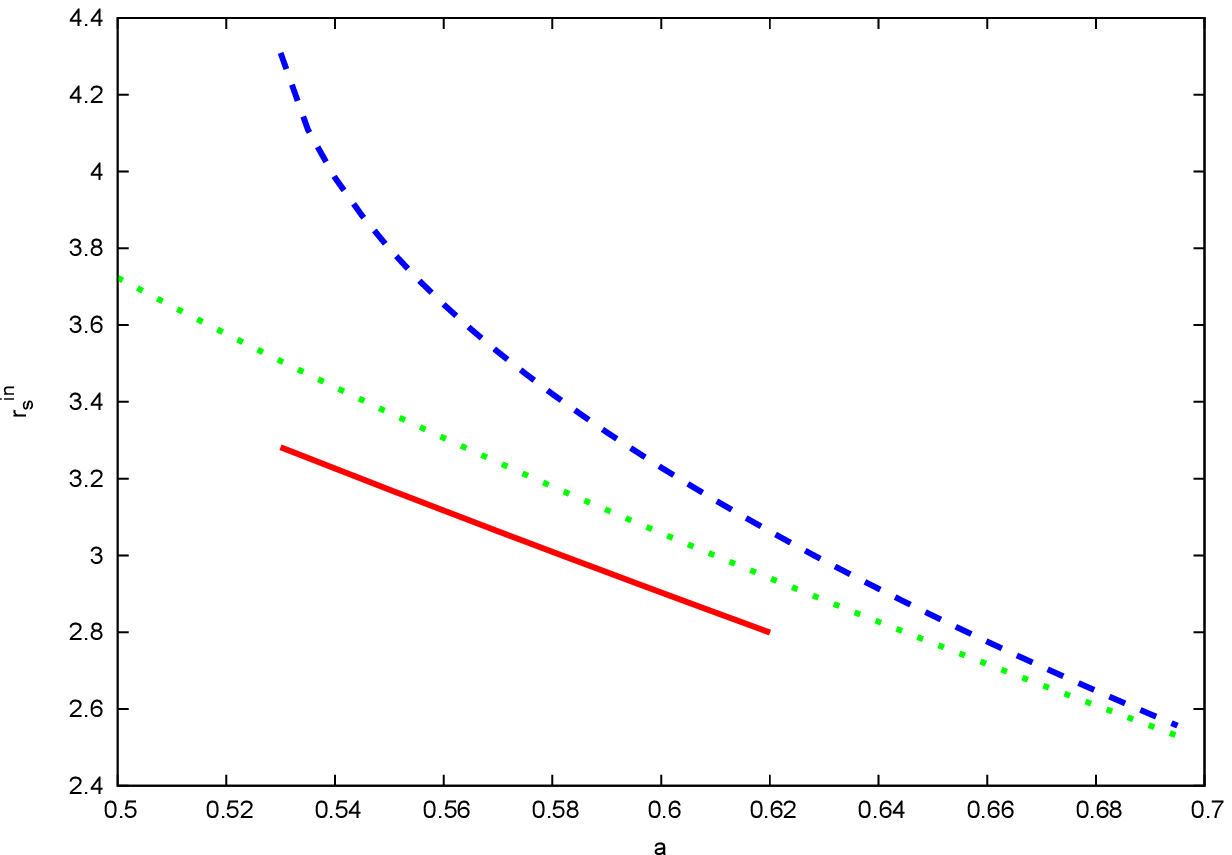} &
\includegraphics[width=0.5\linewidth]{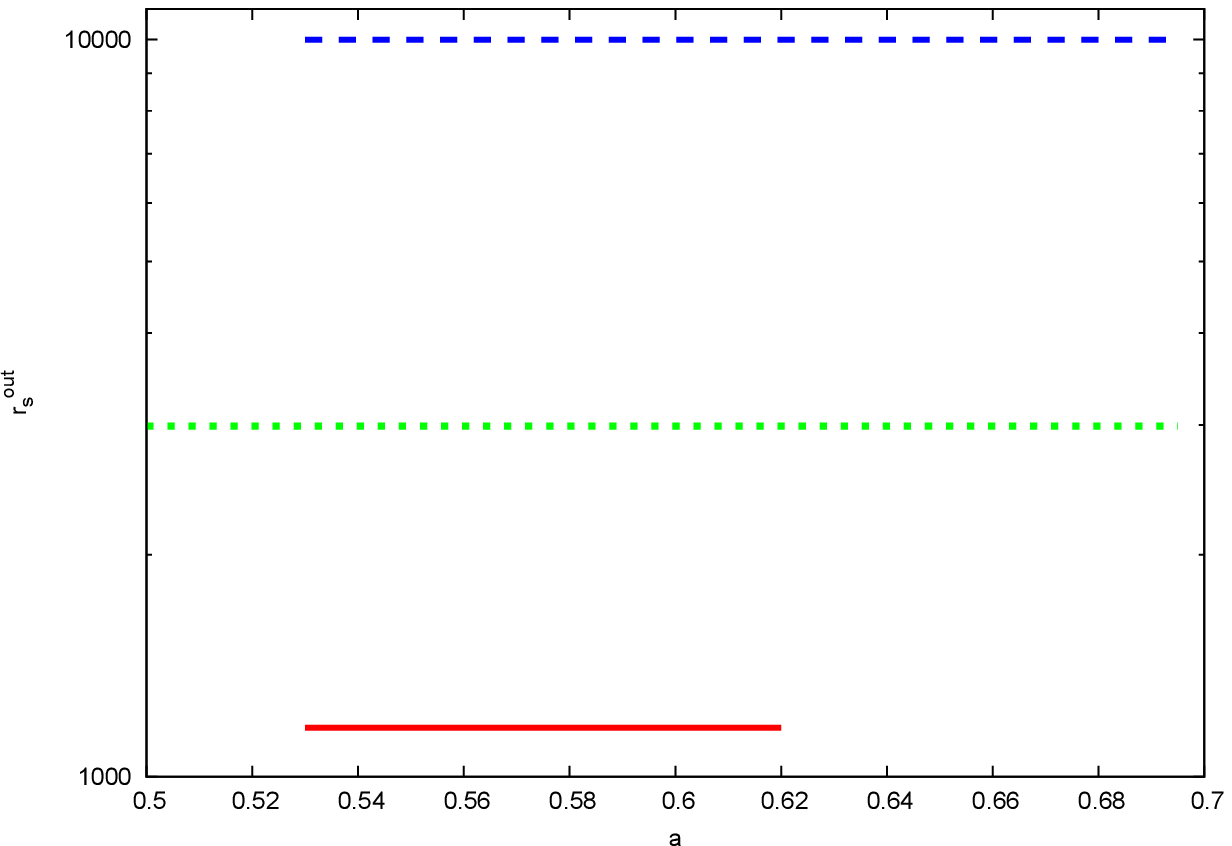}
\end{tabular}
\caption{$r_s^{in} - a$ and $r_s^{out} - a$ (${\cal E}=1.00025,\lambda=3.0,\gamma=4/3$) plots. 
Green dotted line, red solid line and blue dashed line represent CF, VE and CH flows respectively.}
\label{fig2}
\end{figure}

In fig. 2, we plot the locations of the inner acoustic horizon $r_s^{in}$ and the outer acoustic horizons $r_s^{out}$, 
respectively, as functions of $a$, for three different flow geometries. One notes that the influence of the Kerr parameter in changing the values of the 
acoustic horizon locations, is more prominent for the inner horizons. This is somewhat obvious, since the outer type sonic points form at a 
reasonably large distance from the black hole where the space time tends to become Newtonian and the black hole spin does not influence the 
flow properties significantly. \\

\begin{figure}[h!]
\centering
\begin{tabular}{cc}
\includegraphics[width=0.5\linewidth]{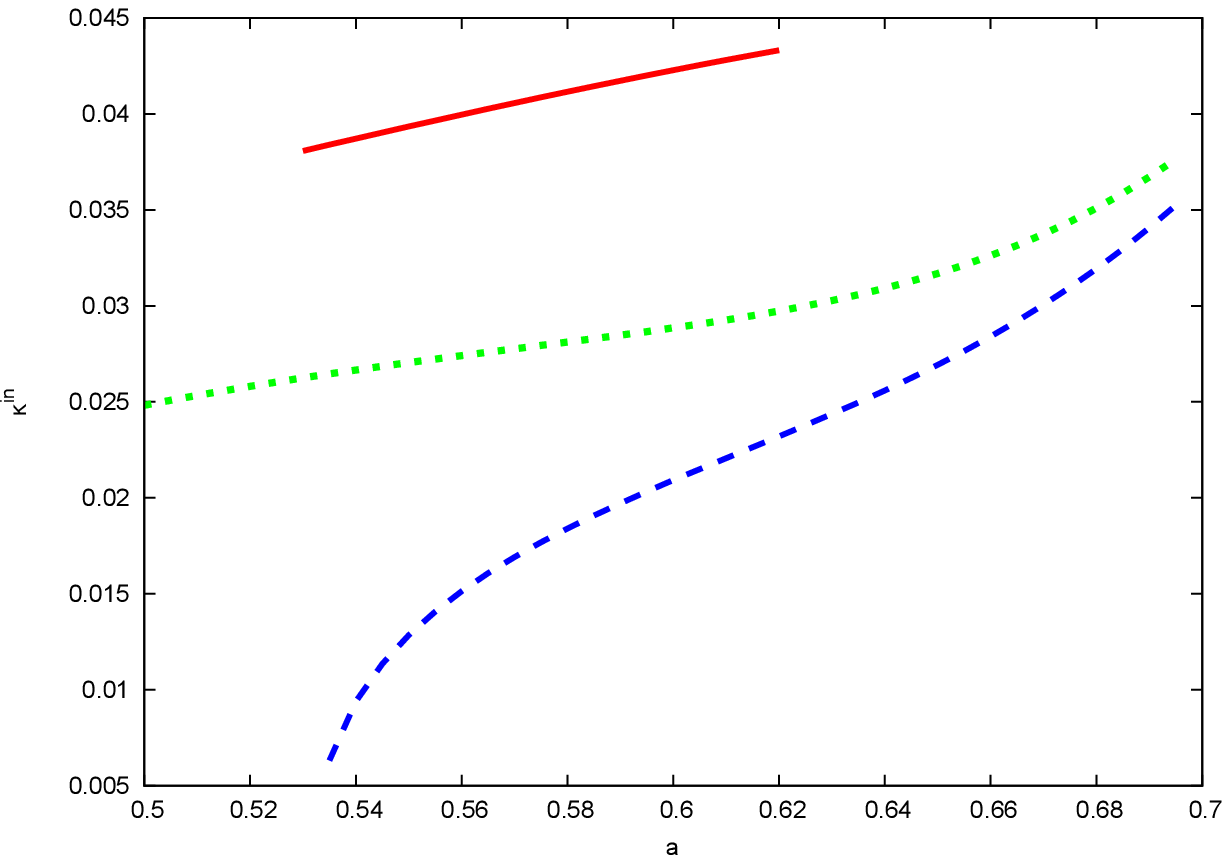} &
\includegraphics[width=0.5\linewidth]{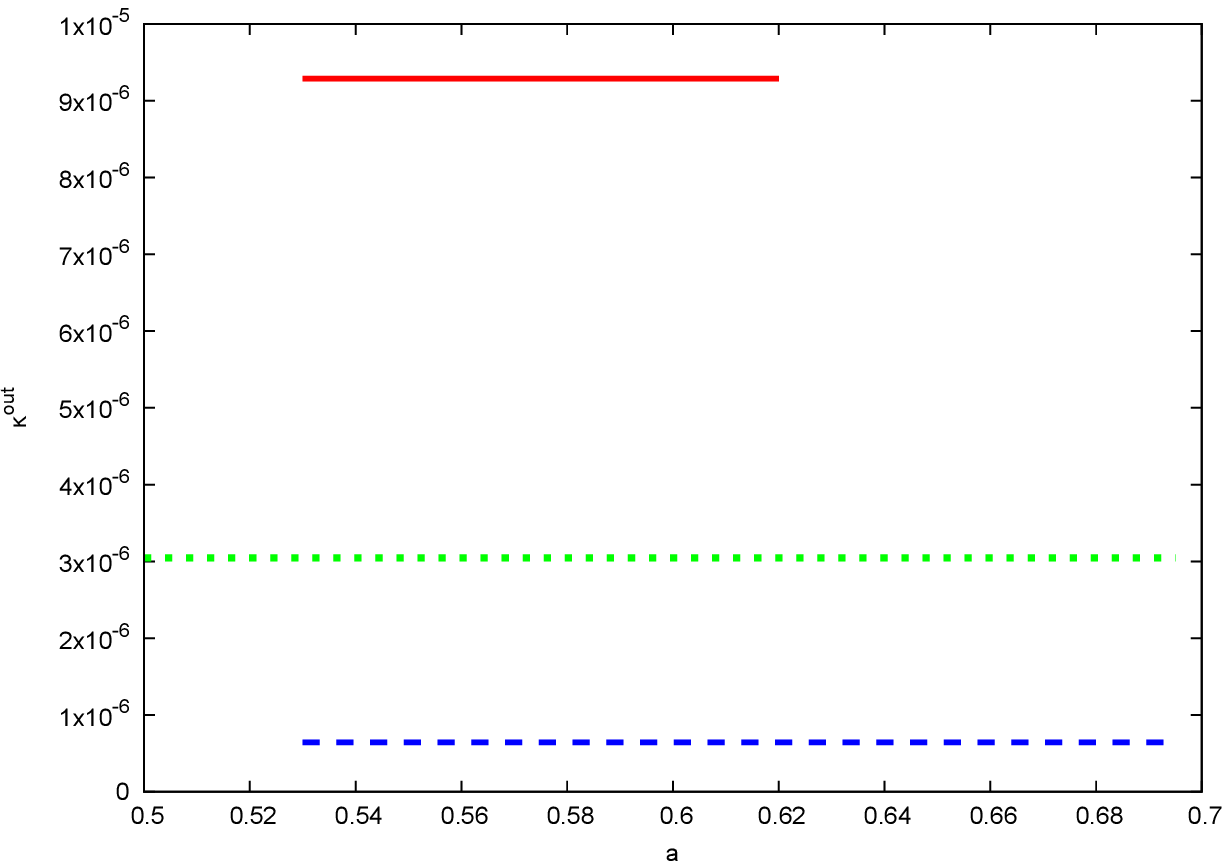}
\end{tabular}
\caption{$\kappa^{in} - a$ and $\kappa^{out} - a$ (${\cal E}=1.00025,\lambda=3.0,\gamma=4/3$) plots. 
Green dotted line, red solid line and blue dashed line represent CF, VE and CH flows respectively.}
\label{fig3}
\end{figure}

In fig. 3, we plot the variation of the acoustic surface gravity (evaluated at the inner and at the outer sonic points, respectively) 
as a function of the black hole spin for three different geometric configurations of matter. As explained in the previous sections, 
it is first necessary to locate the shock forming region in the parameter space which is common to the three different geometries. 
Since, only a small portion of the parameter space is obtained through such procedure, hence the $\kappa - a$ relationship is obtained 
for a limited range of black hole spin, and for all the disc structures, the span of $a$ may not be exactly the same. 
We find that the numerical value of $\kappa^{in}$ is always significantly large compared to the corresponding value of $\kappa^{out}$ for same set of 
values of $\left[{\cal E},\lambda,\gamma,a\right]$. This is, 
once again, somewhat obvious because the $r_s^{in}$ forms closer to the black hole (compared to $r_s^{out}$) and hence at its location gravity becomes 
stronger. It is also observed that the variation of acoustic surface gravity with black hole spin gets significantly influenced by the geometric 
configuration of the flow. From fig. 3, it is observed that \\
\begin{equation}
\kappa_{VE}>\kappa_{CF}>\kappa_{CH}
\label{eqn4}
\end{equation}
for the same value of black hole spin. However, as of now, by plotting the $\kappa - a$ graph for 
multitransonic shocked accretion, we have not been able to cover a reasonably large span of the Kerr parameter. We would like to observe how $\kappa$ varies over the entire astrophysically relevant range of $a$ from -1 to +1. \\
In order to obtain such a plot, we shift our attention from multitransonic shocked accretion to monotransonic accretion. In case of 
monotransonic flow, the three different disc structures overlap almost over the entire range of $a$. This task is absolutely 
essential for us to perform, in order to comment on the global variation of any quantity of our interest (in this case, $\kappa$) with black hole spin. 
It is also interesting to note that such variation covers both prograde and retrograde flows, and thus any asymmetry in the behaviour of the plot 
on positive and negative sides of the $a$-axis points towards a possible instrument to distinguish between the two different kinds of fluid flow 
around the rotating central massive compact objects. 

\begin{figure}[h!]
\centering
\includegraphics[width=\linewidth]{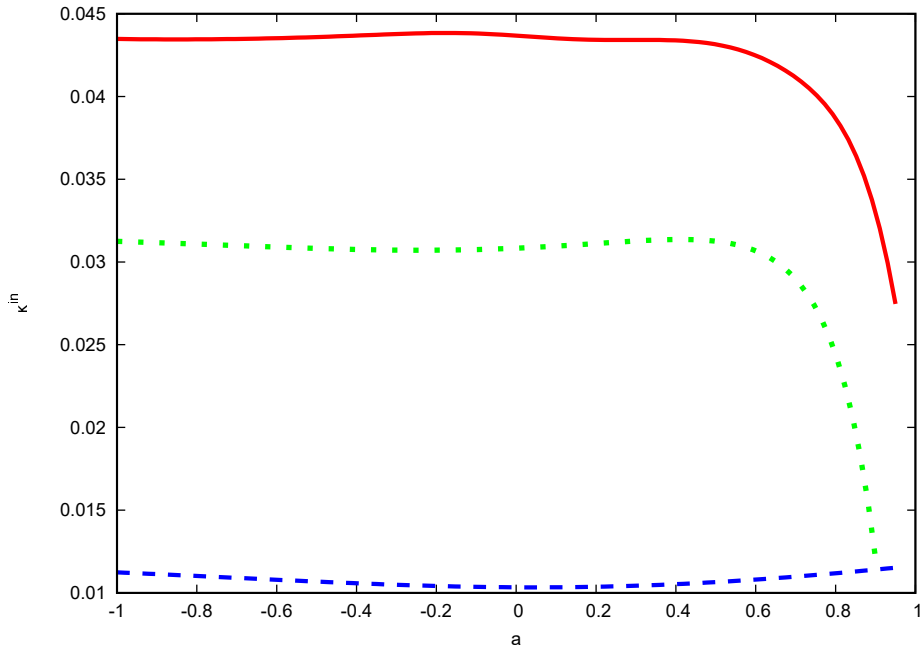}
\caption{$\kappa^{in} - a$ (${\cal E}=1.2,\lambda=2.0,\gamma=4/3$) plots for monotransonic accretion. 
Green dotted line, red solid line and blue dashed line represent CF, VE and CH flows respectively.}
\label{fig4}
\end{figure}

In fig. 4, the variation of $\kappa$ with $a$ has been shown for all three flow geometries 
considering monotransonic accretion. The figure has been obtained for a given set of values for $\left[{\cal E},\lambda,\gamma\right]$ corresponding 
to the stationary monotransonic accretion solutions. It is observed that the constant height disc flow does not have any maxima or point of inflection, 
whereas the other two flow geometries manifest such peaks. This, however, does not necessarily indicate that the constant height disc flow does not 
exhibit maxima in general. This artifact is solely attributed to the choice of $\left[{\cal E},\lambda,\gamma\right]$ particularly for the given figure. 
As a matter of fact, all three flow geometries possess extrema in their dependence of $\kappa$ on spin, however, this figure clearly indicates that there 
is a shift in the position of the peaks with variation of the other flow parameters.

\section{Dependence of $\kappa$ on $a$ for isothermal accretion}

\begin{figure}[h!]
\centering
\includegraphics[width=\linewidth]{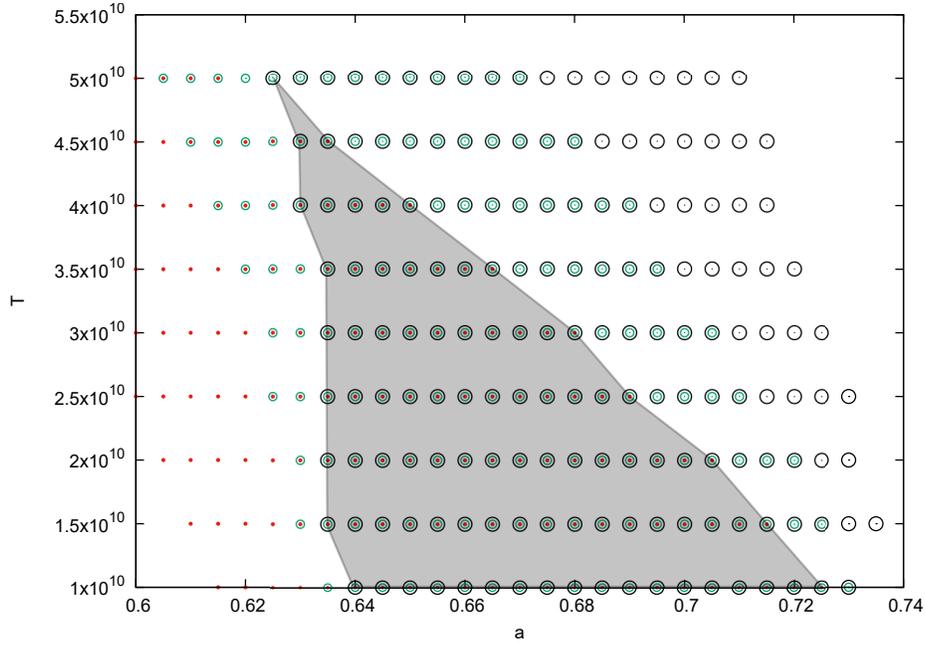}
\caption{$T - a$ ($\lambda=3.0$) plot showing the common shock overlap regions for 
the three disc geometries. $T$ is in units of Kelvin. Green circles, red dots and black circles 
represent CF, VE and CH flows respectively.}
\label{fig5}
\end{figure}

In fig. 5, we show the parameter space spanned by the constant flow temperature ($T$) and the black hole spin to demonstrate the common shock forming 
region for the different geometric configurations of matter. Similar figures can be produced in the $\lambda - a$ parameter space as well. Following the 
same line of discussion in the adiabatic section, one may obtain $\kappa^{in}$, $\kappa^{out}$, $\kappa^{sh}$ in the isothermal case as well, where $
\kappa^{sh}$ cannot be evaluated since the space gradients of the accretion variables diverge at the shock location. 

\begin{figure}[h!]
\centering
\begin{tabular}{cc}
\includegraphics[width=0.5\linewidth]{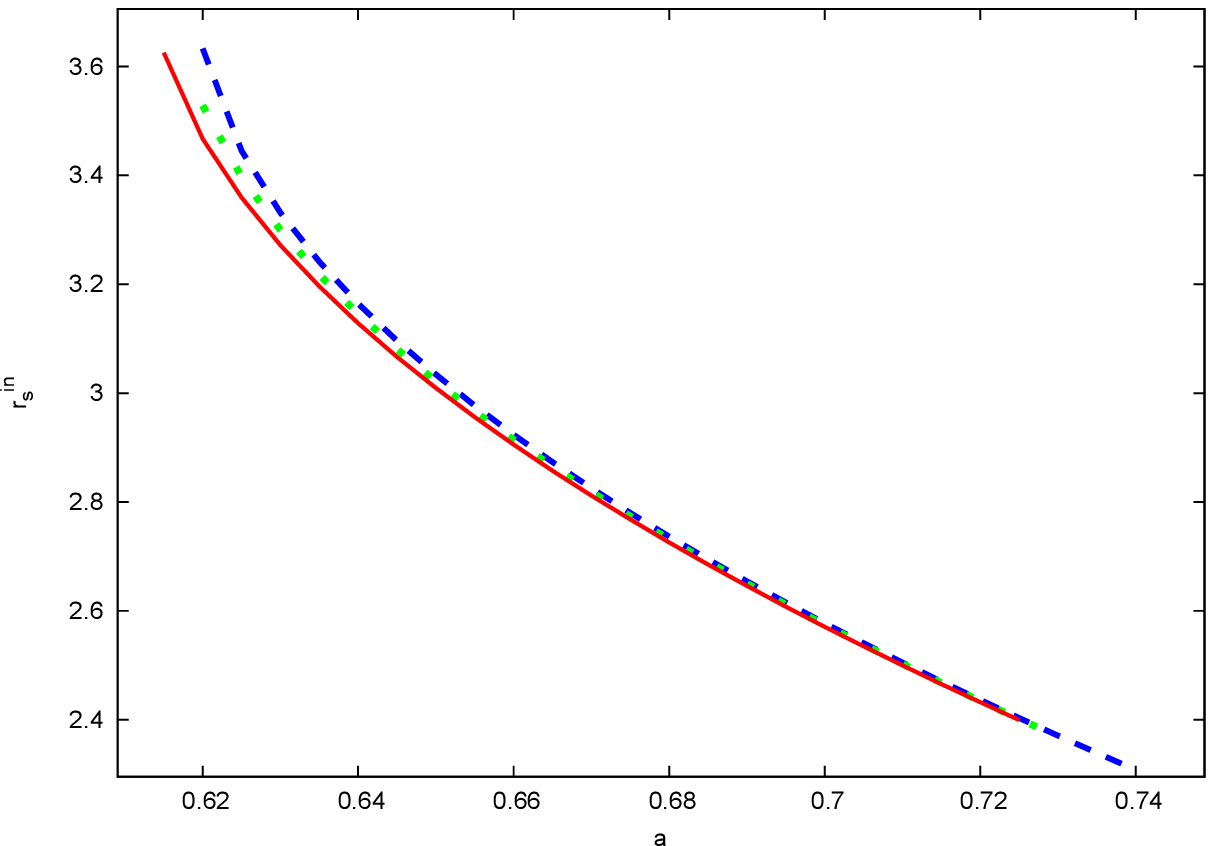} &
\includegraphics[width=0.5\linewidth]{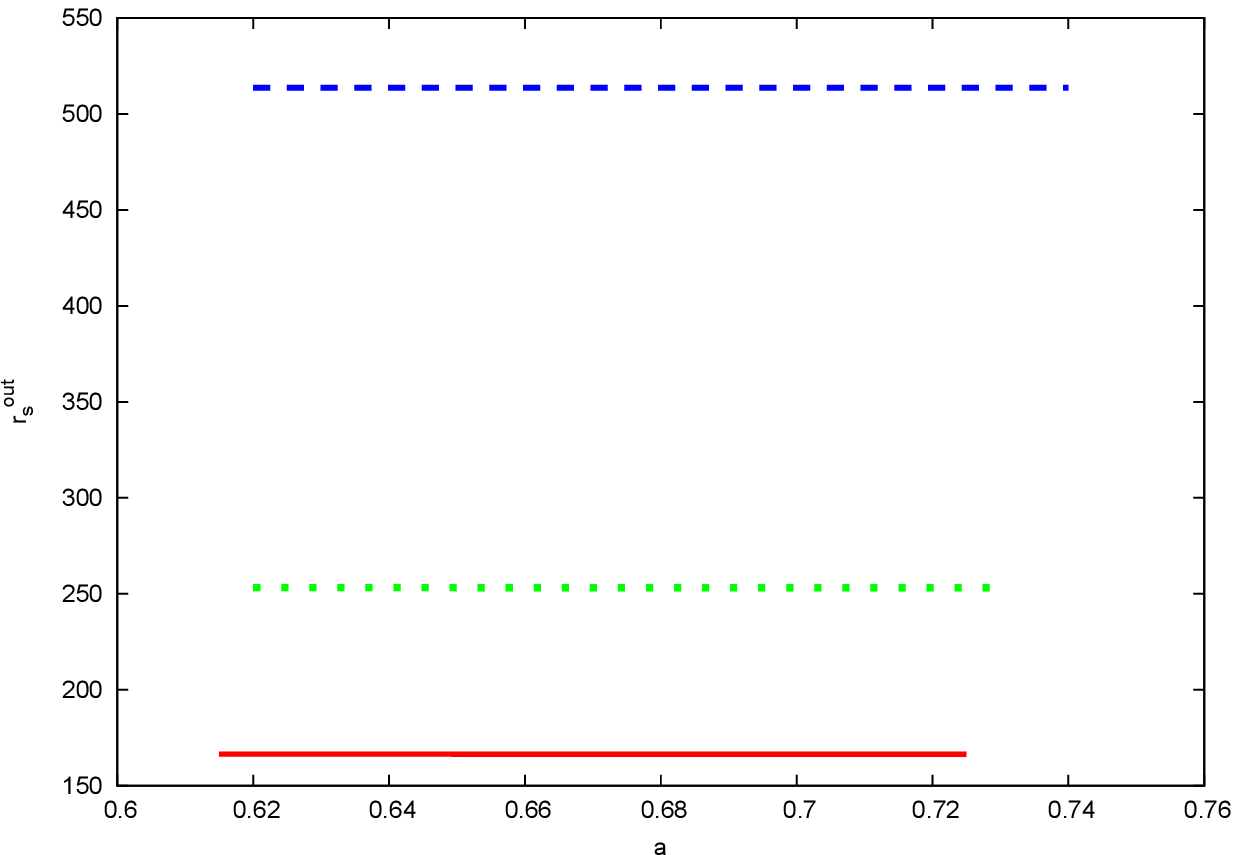}
\end{tabular}
\caption{$r_s^{in} - a$ and $r_s^{out} - a$ ($T=10^{10}$ K, $\lambda=3.0$) plots. 
Green dotted line, red solid line and blue dashed line represent CF, VE and CH flows respectively.}
\label{fig6}
\end{figure}

In fig. 6, spin dependence 
of the location of the inner and the outer horizons are shown. As expected, $a$ has stronger influence in determining the variation of $r_s^{in}$ in 
comparision to the variation of $r_s^{out}$. \\

\begin{figure}[h!]
\centering
\begin{tabular}{cc}
\includegraphics[width=0.5\linewidth]{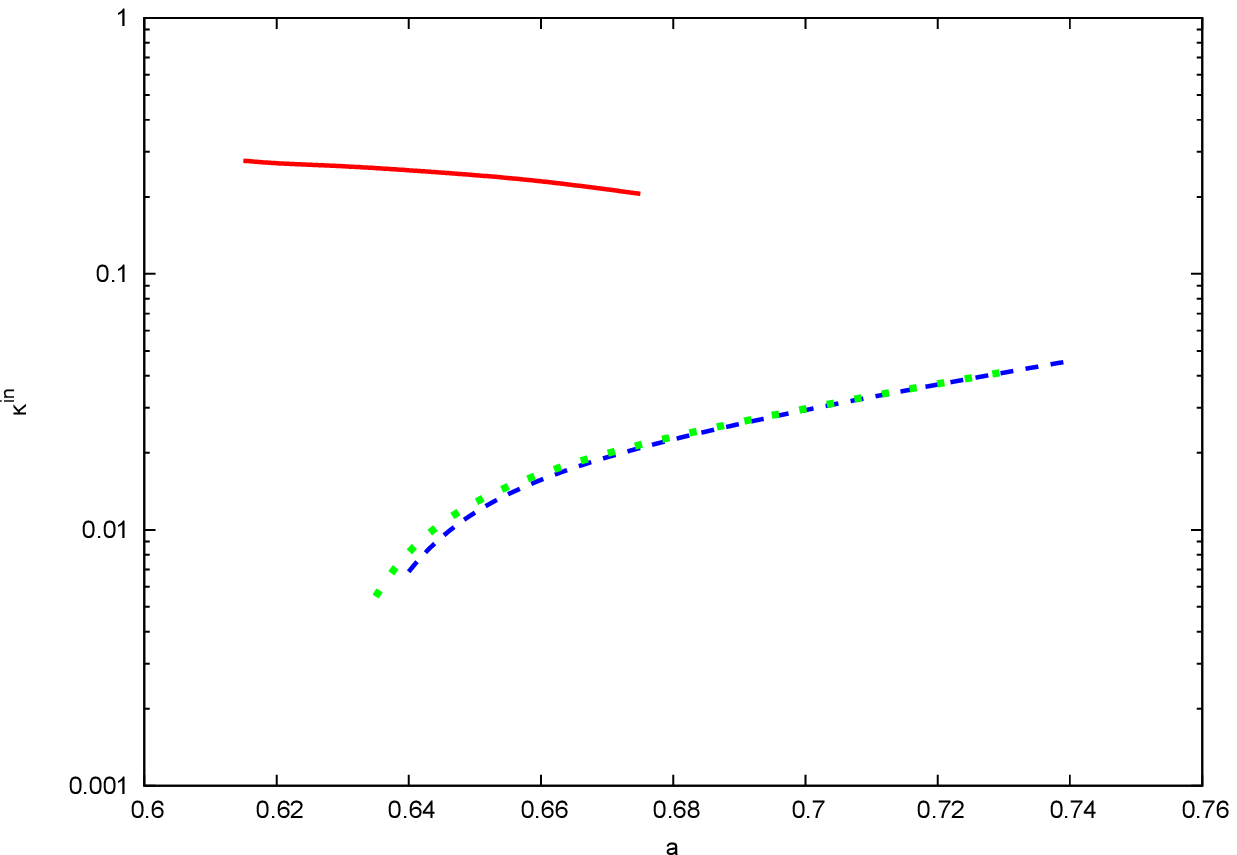} &
\includegraphics[width=0.5\linewidth]{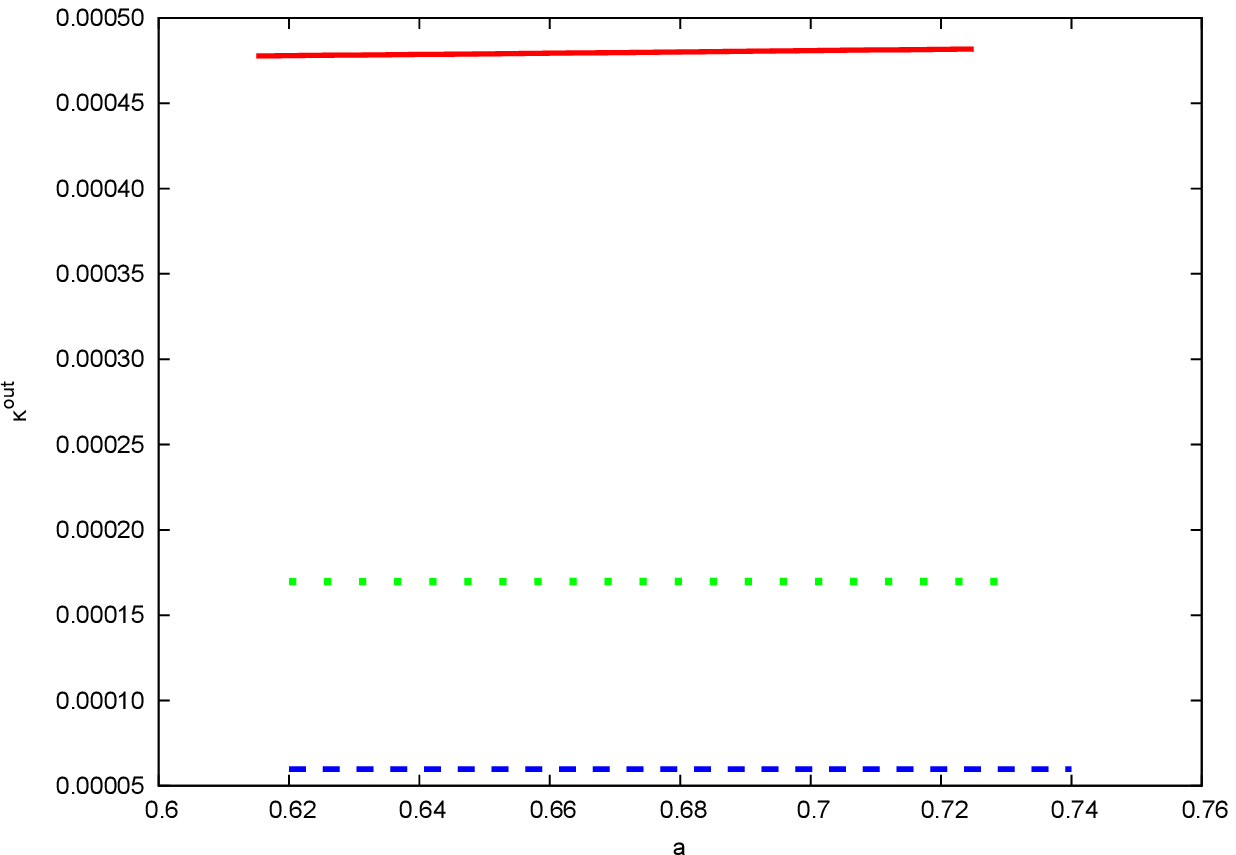}
\end{tabular}
\caption{$\kappa^{in} - a$ and $\kappa^{out} - a$ ($T=10^{10}$ K, $\lambda=3.0$) plots. 
Green dotted line, red solid line and blue dashed line represent CF, VE and CH flows respectively.}
\label{fig7}
\end{figure}

In fig. 7, we plot the variation of $\kappa^{in}$ and $\kappa^{out}$ with the Kerr parameter. In agreement with the discussions presented in the previous 
sections, $\kappa^{in} >> \kappa^{out}$ for all sets of $\left[T,\lambda,a\right]$ also in the case of isothermal accretion. It is also observed that \\
\begin{equation}
\kappa_{VE} > \kappa_{CF} > \kappa_{CH}
\label{eqn5}
\end{equation}
which is in accord with the observations in the polytropic section. However, the individual trend of variation of $\kappa^{in}$ with $a$ for the 
vertical hydrostatic equilibrium disc is found to be contrary to those of the other two disc models. In this context, we must remember 
that multitransonic shocked accretion solutions offer a very narrow window of spin in which all the disc geometries overlap. Hence, behaviour of 
a physical quantity in the given small range of $a$ may not be quite a reliable representation of the bigger picture. Thus, in order to resolve 
this discrepancy, we need to look at the global variation of acoustic surface gravity over the entire span of Kerr parameter. 

\begin{figure}[h!]
\centering
\includegraphics[width=\linewidth]{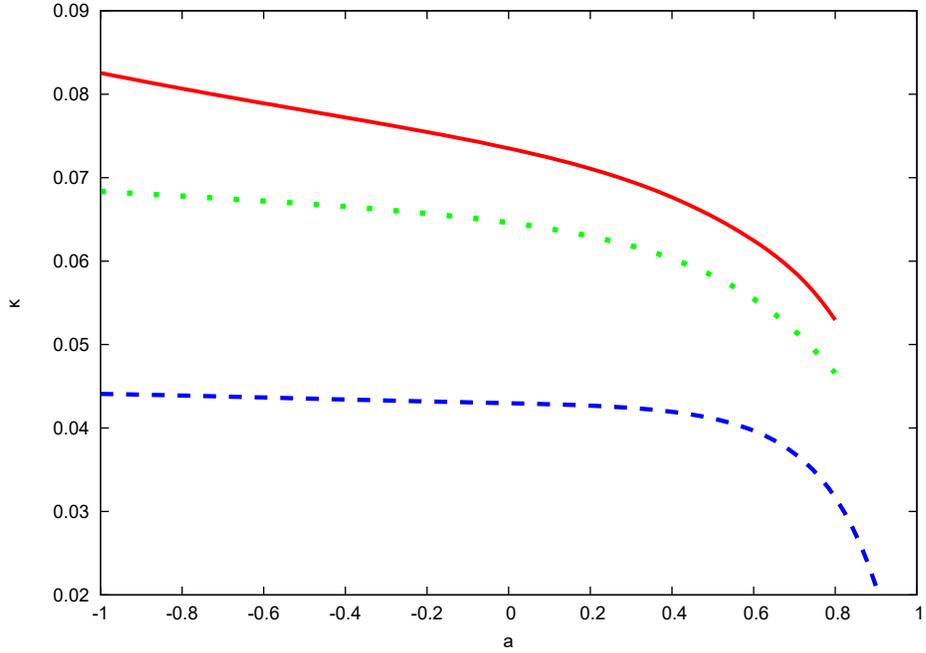}
\caption{$\kappa - a$ ($T=10^{12}$ K, $\lambda=2.0$) plot for monotransonic isothermal accretion. 
$T$ is in units of Kelvin. Green dotted line, red solid line and blue dashed line 
represent CF, VE and CH flows respectively.}
\label{fig8}
\end{figure}

As discussed in the previous section, the only way to have a glimpse of the global dependence of a variable on flow parameters is to study the 
corresponding dependence for monotransonic accretion solutions. Fig. 8 depicts the global variational trend of $\kappa$ over the entire range of black hole 
spin for monotransonic isothermal accretion. It is observed that the trend of the dependence of $\kappa$ on the disc geometry remains the same as for 
polytropic accretion. From the figure, it is clear that a seemingly anti-correlating behaviour amongst the various flow configurations, may actually 
be a manifestation of constrained observations over limited ranges of parameters allowing for a desired class of solutions.

\section{Concluding remarks}

At this stage, it is imperative to reiterate the basic motivation behind the work presented in this manuscript. 
Hawking radiation is universal in the sense that the radiation does not depend on the medium. On the contrary, analogue effects depend on the physical 
properties of matter and such dependence gets reflected through the corresponding dispersion relation, see, e.g., 
\cite{lr12njp,robertson12jpb} for further details. 
In existing works, however, the background space-time has been considered to be flat Minkowskian, and the configuration of the fluid (whether 
classical or quantum) has been assumed to follow certain simplified symmetry, like planar, circular, or spherical, for instance \cite{blv05lrr}. 
Our work is way more 
involved in this context. Firstly, the space-time in which the dynamics of the fluid has been defined, is itself curved since we study the fluid 
motion in general relativistic black hole metric. Also such curved space-time has its variant - it may be of Schwarzschild or of Kerr type - where the 
specific unique characteristic feature of such different background geometries will determine the nature of the embedded acoustic geometry. 
In addition to the geometry of the space-time, the geometry of the fluid itself determines the features of the sonic geometry. In the present 
work, we have shown how the behaviour of the analogue acoustic geometry is determined by the combined effect of the space-time geometry as well as 
the matter geometry. Such an attempt provides rich features of the sonic geometry which has been demonstrated in this work through the 
dependence of the acoustic surface gravity on both the space-time and the matter geometry. We have also shown which disc structure may produce 
maximum value of the analogue temperature (which is a scalar multiple of $\kappa$).

\section*{Acknowledgments}
PT would like to acknowledge the kind hospitality provided by HRI, Allahabad, India, for several 
visits through the $\rm{XII^{th}}$ plan budget of Cosmology and High Energy Astrophysics grant. 
PT intends to acknowledge Md. Arif Shaikh for pointing out crucial corrections in the manuscript. 
The authors also intend to acknowledge the anonymous reviewer for useful comments and suggestions.

\bibliographystyle{plainnat}
\bibliography{paper}

\end{document}